\documentclass[12pt]{article}
\usepackage{wrapfig}
\usepackage{graphicx}
\usepackage{float}
\usepackage[stable]{footmisc}
\usepackage{booktabs}
\usepackage{amsmath}

\newcommand\pubnumber{}
\newcommand\pubdate{\today}

\def\napoli{$^{\rm A}$University of Tsukuba\\$^{\rm B}$High Energy Accelerator Research Organization(KEK)\\ $^{\rm C}$Tohoku University\\}

\def\Title#1{\begin{center} {\Large #1 } \end{center}}
\def\Author#1{\begin{center}{ \sc #1} \end{center}}
\def\Address#1{\begin{center}{ \it #1} \end{center}}

\newcommand\pubblock{\rightline{\begin{tabular}{l} \pubnumber\\
\pubdate \end{tabular}}}
\newenvironment{Abstract}{\begin{quotation} }{\end{quotation}}
\newenvironment{Presented}{\begin{quotation} \begin{center} 
PRESENTED AT\end{center}\bigskip 
\begin{center}\begin{large}}{\end{large}\end{center} \end{quotation}}
\def\Acknowledgements{\bigskip \bigskip \begin{center} \begin{large}
\bf ACKNOWLEDGEMENTS \end{large}\end{center}}

\def\beq{\begin{equation}}
\def\eeq#1{\label{#1}\end{equation}}
\def\eeqn{\end{equation}}

\def\beqa{\begin{eqnarray}}
\def\eeqa#1{\label{#1}\end{eqnarray}}
\def\eeqan{\end{eqnarray}}

\let\bar=\overbar

\def\Dslash{\not{\hbox{\kern-4pt $D$}}}
\def\dslash{\not{\hbox{\kern-2pt $\del$}}}

\def\msb{{\bar{\ssstyle M \kern -1pt S}}}

\begin{document}
\begin{titlepage}
\pubblock

\vfill
\Title{TID-Effect Compensation and Sensor-Circuit Cross-Talk Suppression in Double-SOI Devices}
\vfill
\Author{ Shunsuke Honda$^{\rm A}$, Kazuhiko Hara$^{\rm A}$, Daisuke Sekigawa$^{\rm A}$, Bipin Subedi$^{\rm A}$, Mari Asano$^{\rm A}$, Naoshi Tobita$^{\rm A}$, Wataru Aoyagi$^{\rm A}$, Yasuo Arai$^{\rm B}$, Akimasa Ishikawa$^{\rm C}$, Yoshimasa Ono$^{\rm C}$, Itaru Ushiki$^{\rm C}$, SOI Collaboration }
\Address{\napoli}
\vfill
\begin{Abstract}
We are developing double silicon-on-insulator (DSOI) pixel sensors for various applications such as for high-energy experiments. The performance of DSOI devices has been evaluated including total ionization damage (TID) effect compensation in transistors using a test-element-group (TEG) up to 2~MGy and in integration-type sensors up to 100~kGy. In this article, successful TID compensation in a pixel-ASD-readout-circuit is shown up to 100~kGy for the application of DSOI to counting-type sensors. The cross-talk suppression in DSOI is being evaluated. These results encourage us that DSOI sensors are applicable to future high-energy experiments such as the BELLE-II experiment or the ILC experiment. 
\end{Abstract}
\vfill
\begin{Presented}
International Workshop on SOI Pixel Detector (SOIPIX2015), Tohoku University, Sendai, Japan, 3-6, June, 2015.
\end{Presented}
\vfill
\end{titlepage}
\def\thefootnote{\fnsymbol{footnote}}
\setcounter{footnote}{0}

\section{Introduction}
\begin{figure}[!htbp]
\begin{center}
\includegraphics[width=.6\textwidth]{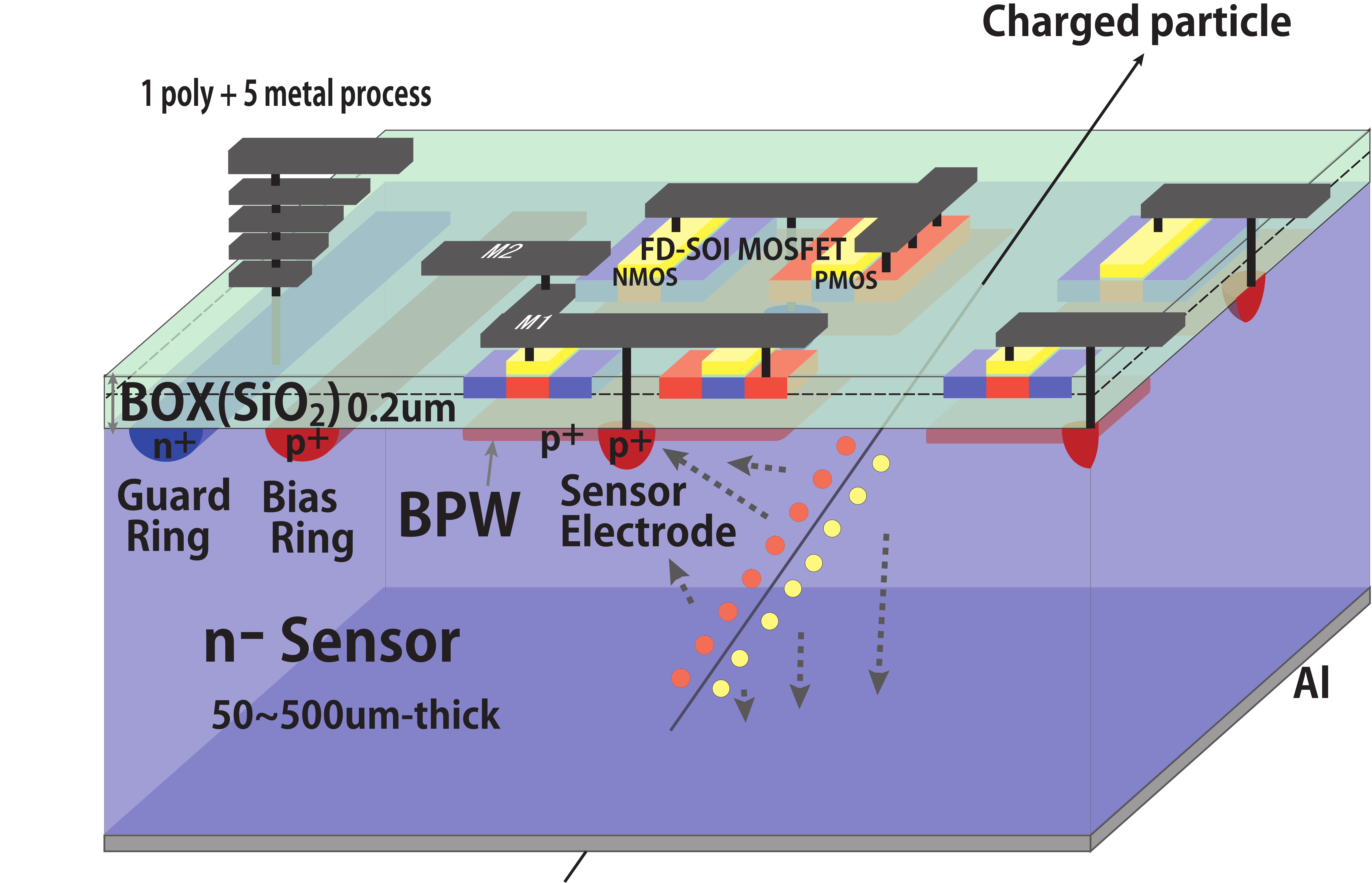}
\end{center}
\caption{Schematics of SOI monolithic pixel device. Two silicon substrates with different resistivities are bonded interleaved with 200-nm thick buried oxide layer (BOX) with each resistivity optimized depending on the sensor part or circuit part. The pixel implantation is made after removing the top silicon and BOX layers. The transistors are inter-connected via aluminum metal placed on top (five metal layers are available). The detector bias is applied between the backplane and the front contact (bias ring). The Buried P-Well (BPW) can suppress the back-gate effect on the front transistor circuits.}
\label{fig:soi}
\end{figure}

Monolithic pixel devices utilizing a 0.2 $\rm\mu$m fully depleted silicon-on-insulator (FD-SOI) technology have been intensively explored for various applications \cite{bib:hep}\cite{bib:xr}\cite{bib:sophias}. 
The main technological features in our development are use of "bonded"-wafers fabricated by "SmartCut$^{\rm TM}$" technology by SOITEC Co.\cite{bib:soitec} and employment of commercially reliable 0.2-$\mu$m process by Lapis Semiconductor\cite{bib:lapis}. The monolithic feature is illustrated in Fig.~\ref{fig:soi} and technology characteristics of the Lapis SOI pixel process are summarized in Table~\ref{tab:lapis}. 

\begin{table}[htb]
\caption{Lapis FD-SOI process}
\label{tab:lapis}
\begin{center}
\setlength{\tabcolsep}{4pt}
\footnotesize
\begin{tabular}{@{}cl@{}} \toprule
Process & 0.2$\mu$m low leakage fully-depleted SOI CMOS \\ \midrule
& 1poly+5metal layers, MIM(1.5 $\mu$F/$\mu$m$^2$), DMOS \\
& Core(I/O) voltage = 1.8(3.3) V \\ \hline
SOI Wafer & Diameter: 200 mm$\phi$ \\
& Top Si: CZ (p $\sim$18 $\Omega$cm), 40 nm thick \\
& Buried oxide: 200 nm thick \\
& Handle wafer: CZ (n $\sim$0.7 k$\Omega$cm), FZ (n $\sim$7 k$\Omega$cm / p $\sim$25 k$\Omega$cm) \\
& Double-SOI available \\ \hline
Backside & Mechanical grind (down to 200 $\mu$m), chemical etching, \\
& implant, laser annealing, Al plating. \\
& Further thinning to 50,100 $\mu$m available. \\ \bottomrule
\end{tabular}
\end{center}
\end{table}

Double-SOI (DSOI) has been introduced as a solution to issues arising in application of SOI pixel sensors to high energy experiments. DSOI has an additional Si-layer (SOI2) in BOX, where the independent voltage can be applied. As the most main issue, the total ionization dose (TID) effects \cite{bib:tid} are rather substantial in SOI since each SOI transistor is fully enclosed in oxide layers. With accumulating radiation, generated holes are trapped in BOX affecting the operation of SOI circuits \cite{bib:tkb1}\cite{bib:tkb2}\cite{bib:tkb3}. The TID compensation in DSOI has been evaluated on transistor operation using test-element-group (TEG) chips up to 2~MGy and on integration-type sensors up to 100~kGy. These studies \cite{bib:honda1}\cite{bib:honda2} conclude that the TID effects in SOI devices can be diminished sufficiently up to at least 100~kGy by compensation in DSOI, extending the radiation tolerance nearly two orders in the radiation dose. Recently we have shown the radiation tolerance can be improved further by adjusting the process parameters \cite{bib:kurachi}.  

In this article, TID compensation in a DSOI pixel-ASD-readout-circuit is detailed, which is demonstrated successfully up to 100~kGy. 
The DSOI is also expected to be effective in suppression of cross-talk between sensor nodes and near-by SOI circuit, which is typical in devices with a thin BOX layer. The cross-talk suppression in DSOI is also being addressed.

\section{TID Compensation in Pixel ASD-Readout}

\begin{figure}[htbp]
\begin{center}
\includegraphics[width=\textwidth]{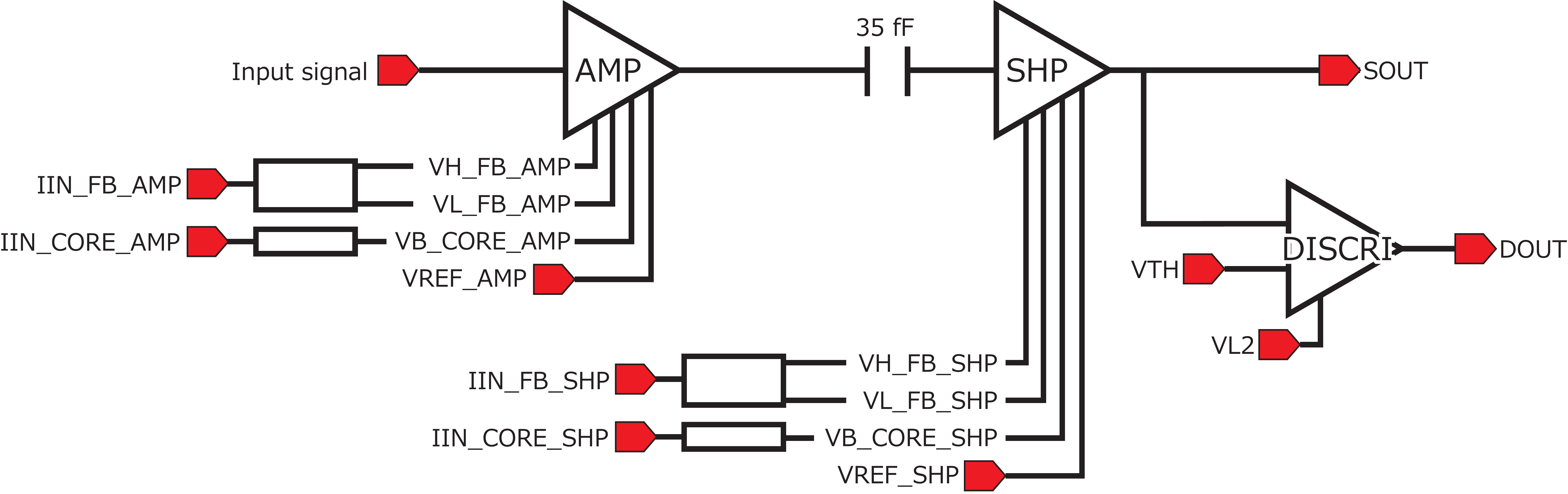}
\end{center}
\caption{Schematics of the ASD-TEG. The output signals from the shaper and the discriminator are measured individually. There are eight inputs (voltage or current) to tune the ASD performance. The four input currents are to generate six appropriate voltages through current mirror circuits (shown in square boxes). The total of ten voltages are explained in testing of each component of ASD (Figs.~\ref{fig:asdtegamp}$\sim$\ref{fig:asdtegdiscri}). 
In addition, there are other ASD circuits where individual components can be tested independently. The parameters are identical as those shown in the figure.}
\label{fig:asdteg}
\end{figure}

The previous studies shown in \cite{bib:honda2} demonstrated the TID compensation in overall functionality of an integration-type pixel sensor. In order to investigate further the detailed TID compensation, we evaluated TID compensation in a TEG device including amplifier-shaper-discriminator (ASD) readout circuit with the response from each stage monitored out. Such a device is for a counting-type detector applicable in high-energy experiments. The schematics of the TEG circuit is shown in Fig.~\ref{fig:asdteg}. The circuit is designed with a readout frequency less than 1~MHz and applicable to high-energy physics experiments such as BELLE-II and ILC. 

The ASD-TEG chips were irradiated up to 100~kGy with $\rm ^{60}$Co $\rm \gamma$-rays at Takasaki Advanced Radiation Research Institute of JAEA. During the irradiation period of two days, all the chip terminals were grounded with irradiation made at room temperature. The samples were brought within four hours at room temperature, then kept at -20$^\circ$C in a refrigerator except during the measurement.

\begin{figure}[htbp]
\begin{minipage}{0.48\hsize}
\begin{center}
\includegraphics[width=1.\textwidth]{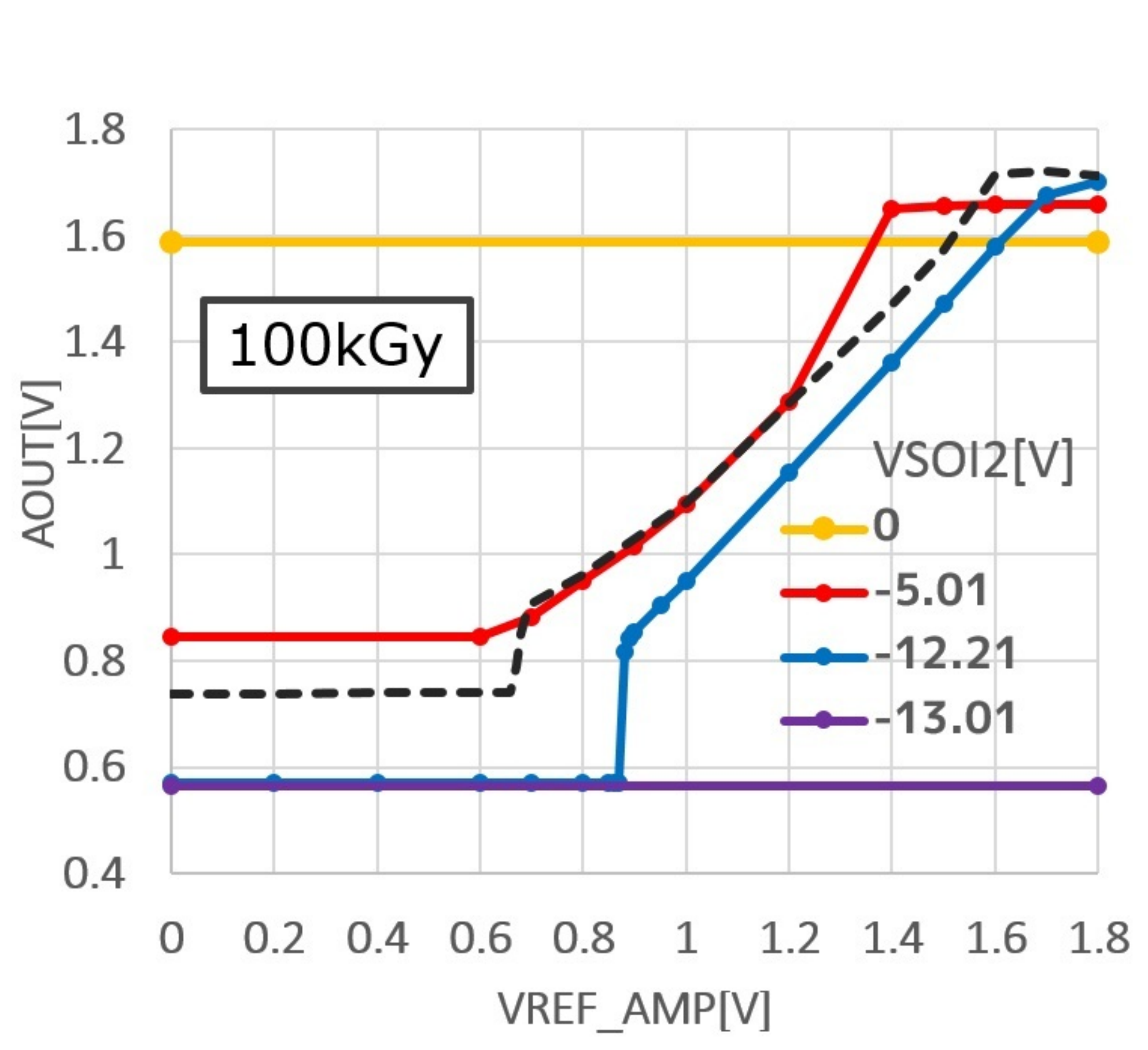}
\end{center}
\caption{The response-curve to VREF in the amplifier after 100kGy. Although the response is lost at V$\rm _{SOI2}$ = 0~V, the response is recovered by applying V$\rm _{SOI2}$ in an appropriate range about from -5~V to -12~V. }
\label{fig:ampdc}
\end{minipage}
\hspace{0.04\hsize}
\begin{minipage}{0.48\hsize}
\begin{center}
\includegraphics[width=1.\textwidth]{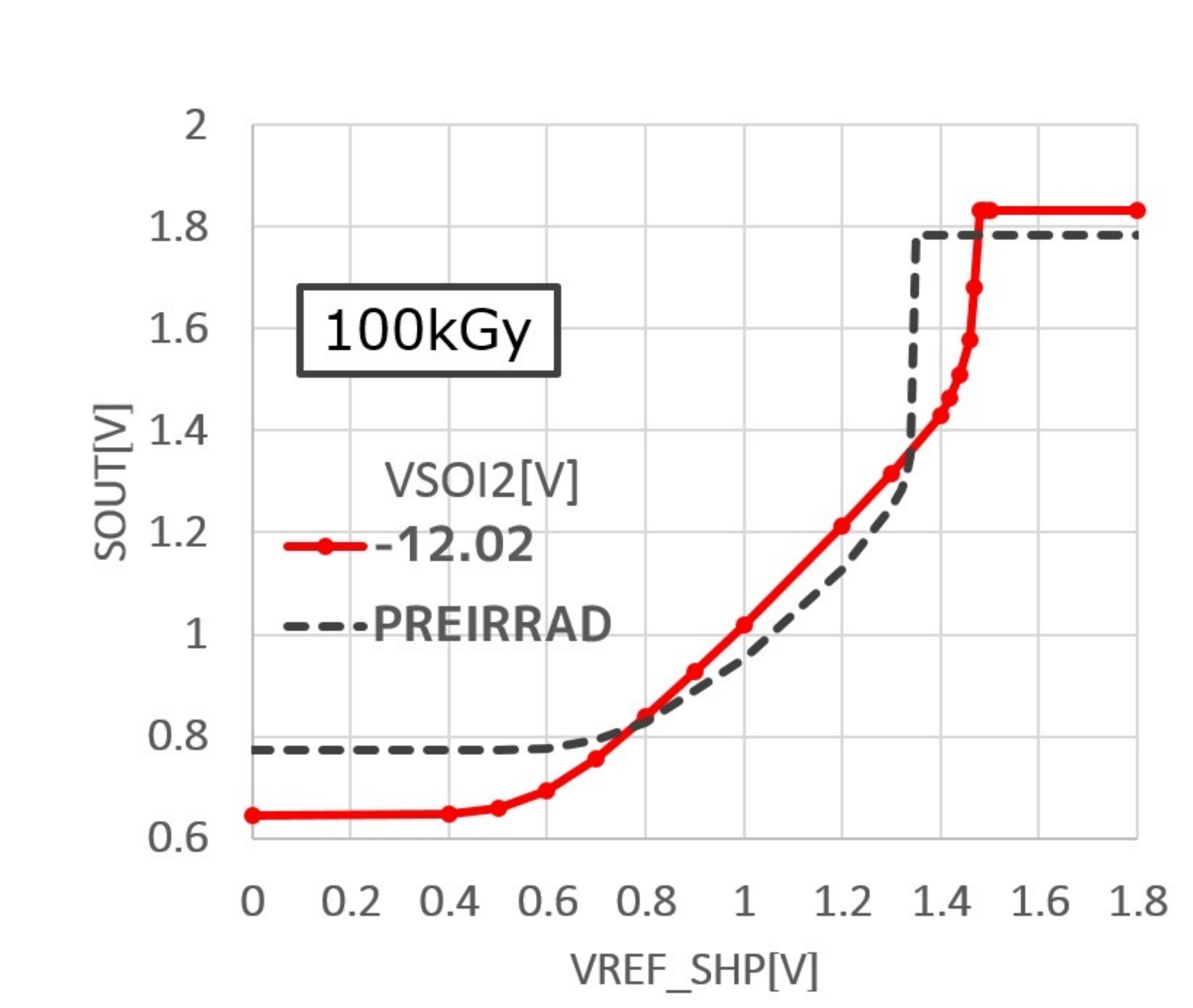}
\end{center}
\caption{The response-curve to VREF in the shaper after 100~kGy. Although the response is lost at V$\rm _{SOI2}$ = 0~V, the response is recovered by applying V$\rm _{SOI2}$. The optimum V$\rm _{SOI2}$ range is narrower in the shaper because larger drive currents are required and the control is more sensitive to V$\rm _{SOI2}$. }
\label{fig:shpdc}
\end{minipage}
\end{figure}

At first, the functionality of each of the amplifier and the shaper, measured with respect to VREF, is shown in Fig.~\ref{fig:ampdc} and Fig.~\ref{fig:shpdc}. 
After the irradiation, the base-line of the output increased and exceeded beyond the dynamic range, disabling evaluation of the response curve at V$\rm _{SOI2}$ = 0~V. But similar to the case as reported in \cite{bib:honda2}, applying a proper V$\rm _{SOI2}$ can recover the lost functionality to VREF. Full recovery is not realized because a single common V$\rm _{SOI2}$ is employed in the TEG chip where various types of transistors are having  different preferred V$\rm _{SOI2}$s depending on transistor parameters. Although much better recovery should be possible by employing multiple V$\rm _{SOI2}$s, we determine here first appropriate ranges of V$\rm _{SOI2}$ depending on the radiation dose to recover both the amplifier and shaper functionalities in wider dynamic range. 

\begin{figure}[htbp]
\begin{center}
\includegraphics[width=0.8\textwidth]{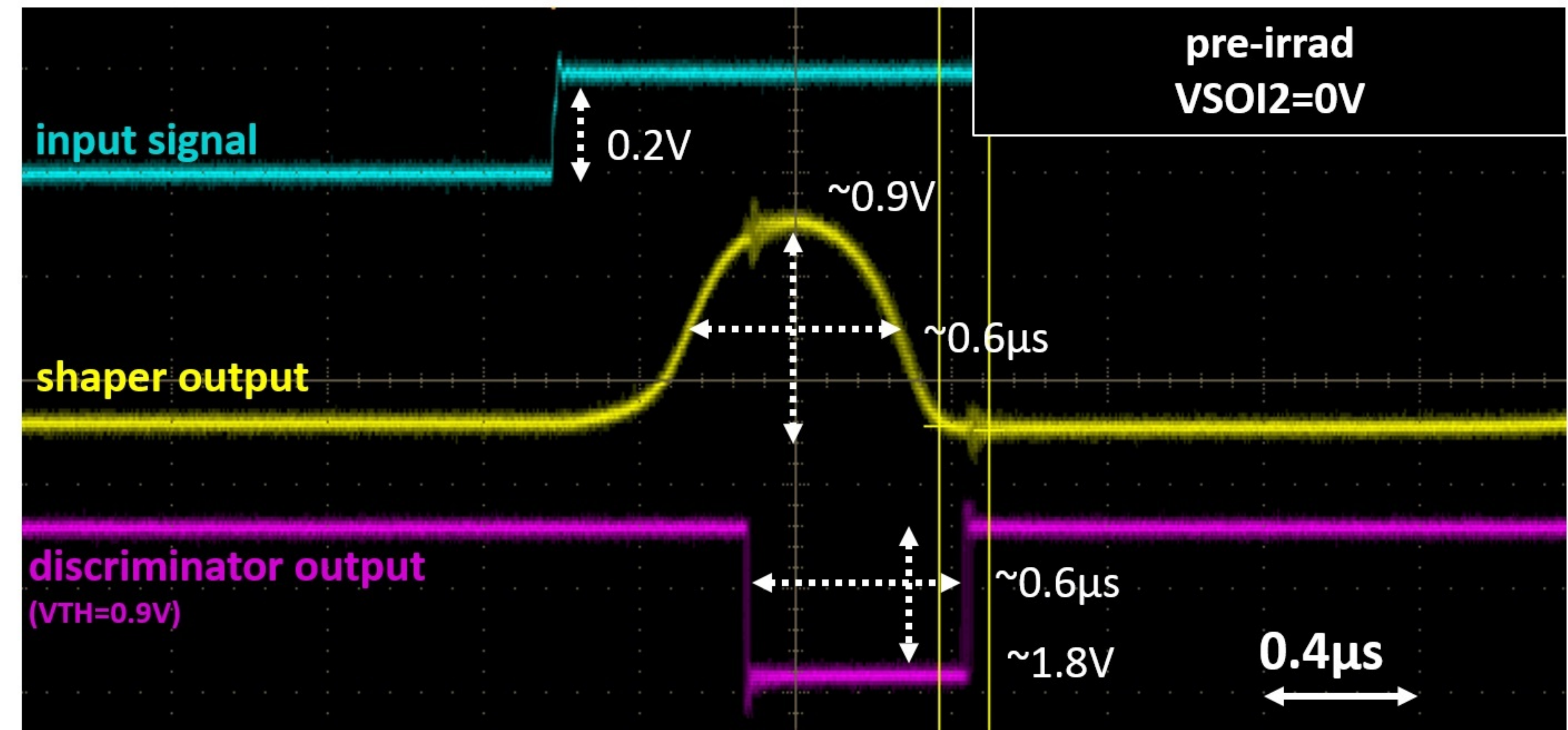}
\end{center}
\caption{The time response of the shaper and the discriminator signals for a pre-irradiation sample. The ASD control currents and voltages are optimized. The discriminator threshold is 0.9~V.}
\label{fig:asdacpi}
\begin{center}
\includegraphics[width=0.8\textwidth]{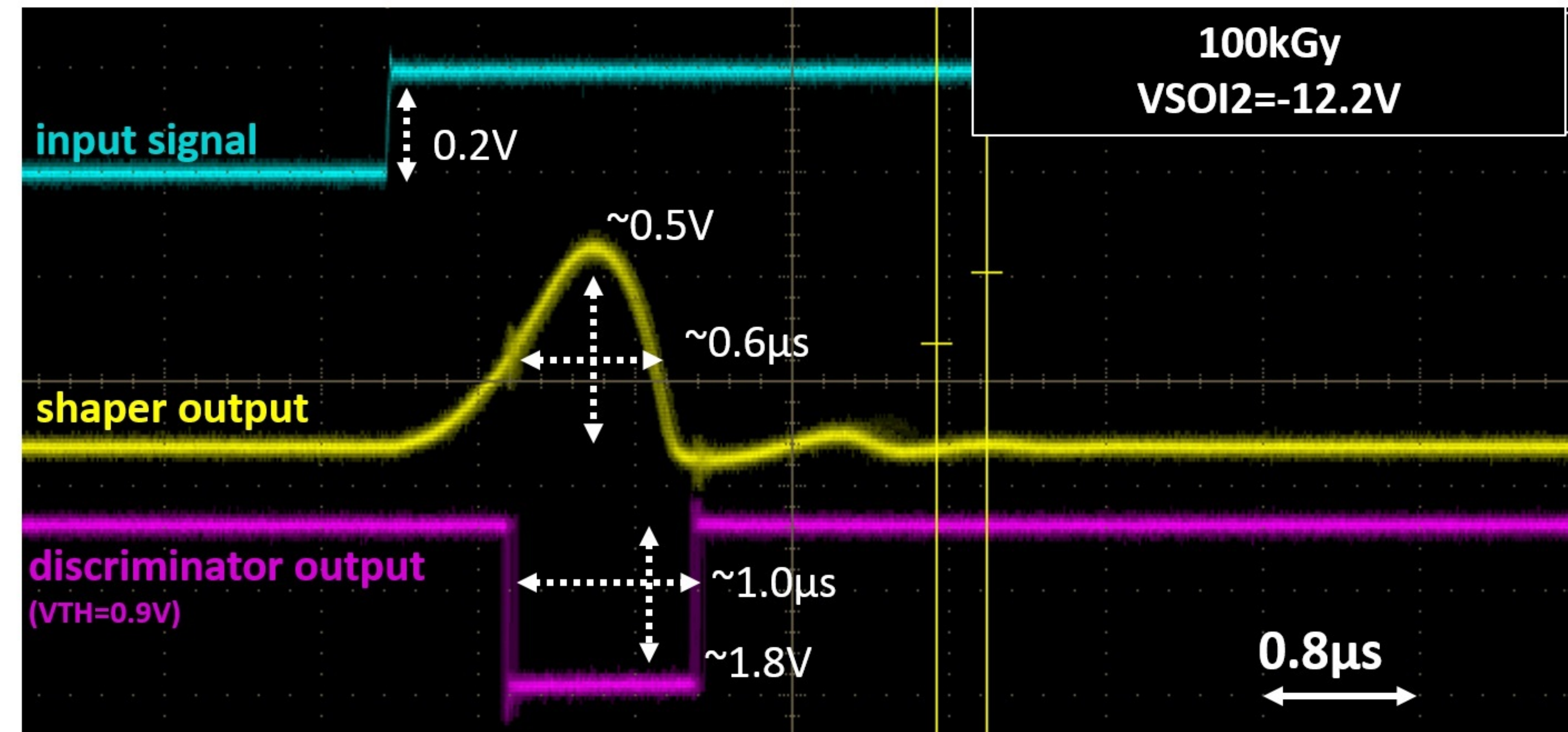}
\end{center}
\caption{The response signals from the shaper and the discriminator after 100~kGy radiation. V$\rm _{SOI2}$ = 12.2~V. The ASD control currents and voltages are optimized. The discriminator is 0.9~V, same as for pre-irradiation.}
\label{fig:asdac100}
\end{figure}

\begin{figure}[htbp]
\begin{center}
\includegraphics[width=0.7\textwidth]{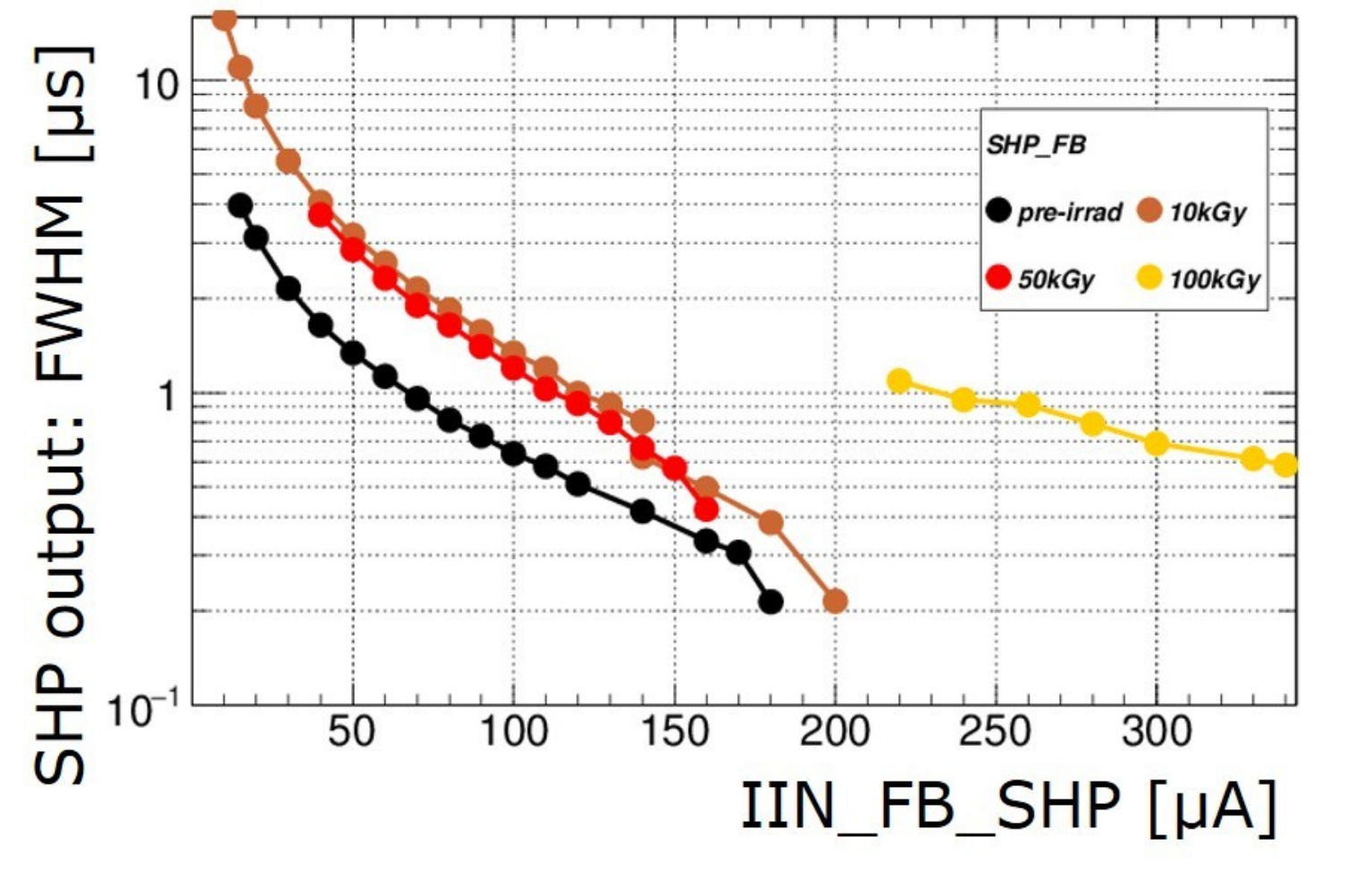}
\end{center}
\caption{The FWHM time-widths of the shaper signal output as function of radiation dose with an optimized V$\rm _{SOI2}$s. V$\rm _{SOI2}$ = -6.0~V at 10~kGy, -9.5~V at 50~kGy, and -12.2~V at 100~kGy.}
\label{fig:shpacfwhm}
\end{figure}

The ASD response to pulse-signal was evaluated by injecting step pulses into the TEG chip through a 1~pF capacitor. The response measured at pre-irradiation is shown in Fig.~\ref{fig:asdacpi}. The shaper and the discriminator outputs have FWHM time-widths of less than 1 $\rm\mu$sec with the appropriate $\rm IIN\_FB\_AMP(SHP)$. After 100~kGy radiation, the response was also obtained with the optimized V$\rm _{SOI2}$ = 12.2~V (Fig.~\ref{fig:asdac100}). For evaluatioing further dependence on the input current ($\rm IIN\_FB\_SHP$), the FWHM time-widths of the output shaper signal are shown in Fig.~\ref{fig:shpacfwhm}. Even after 100~kGy, we can maintain the original 1~$\rm\mu$sec FWHM although a larger input current ($\rm IIN\_FB\_SHP$) is required.

\begin{figure}[htbp]
\begin{minipage}{0.48\hsize}
\begin{center}
\includegraphics[width=1.\textwidth]{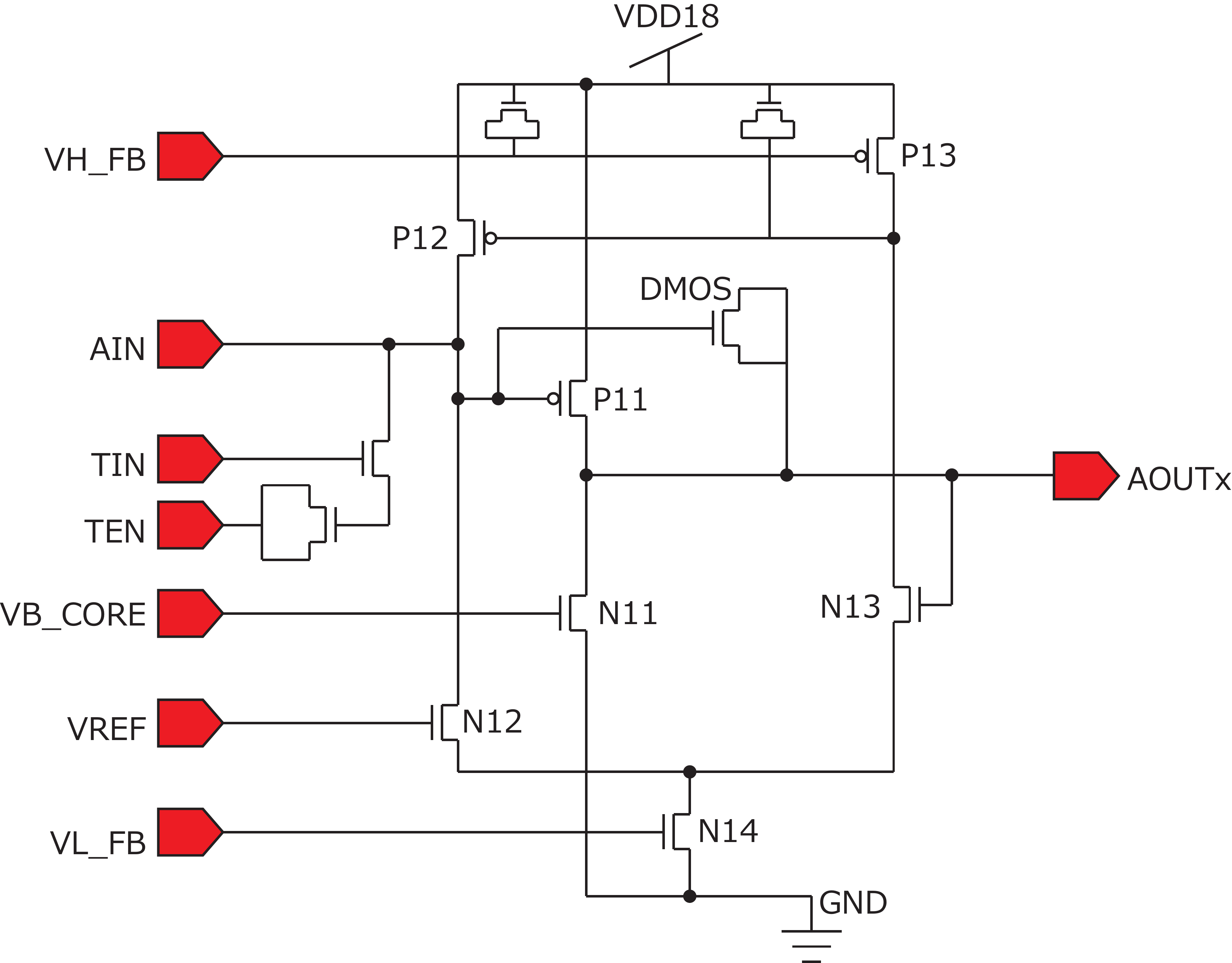}
\end{center}
\caption{The schematics of the amplifier in ASD-TEG. The amplifier is based on a common source circuit with Krummenacher-scheme feedback and noise-suppression \cite{bib:krummenacher}. The input signal is injected through AIN. VREF voltage determines the input and output offsets. $\rm VB\_CORE$ voltage generated from $\rm IIN\_CORE\_AMP$ adjusts the amplifier gain. $\rm VH\_FB$ and $\rm VL\_FB$ generated from $\rm IIN\_FB\_AMP$ determines the feedback current and adjusts the time-width of the output signal.}
\label{fig:asdtegamp}
\end{minipage}
\hspace{0.04\hsize}
\begin{minipage}{0.48\hsize}
\begin{center}
\includegraphics[width=1.\textwidth]{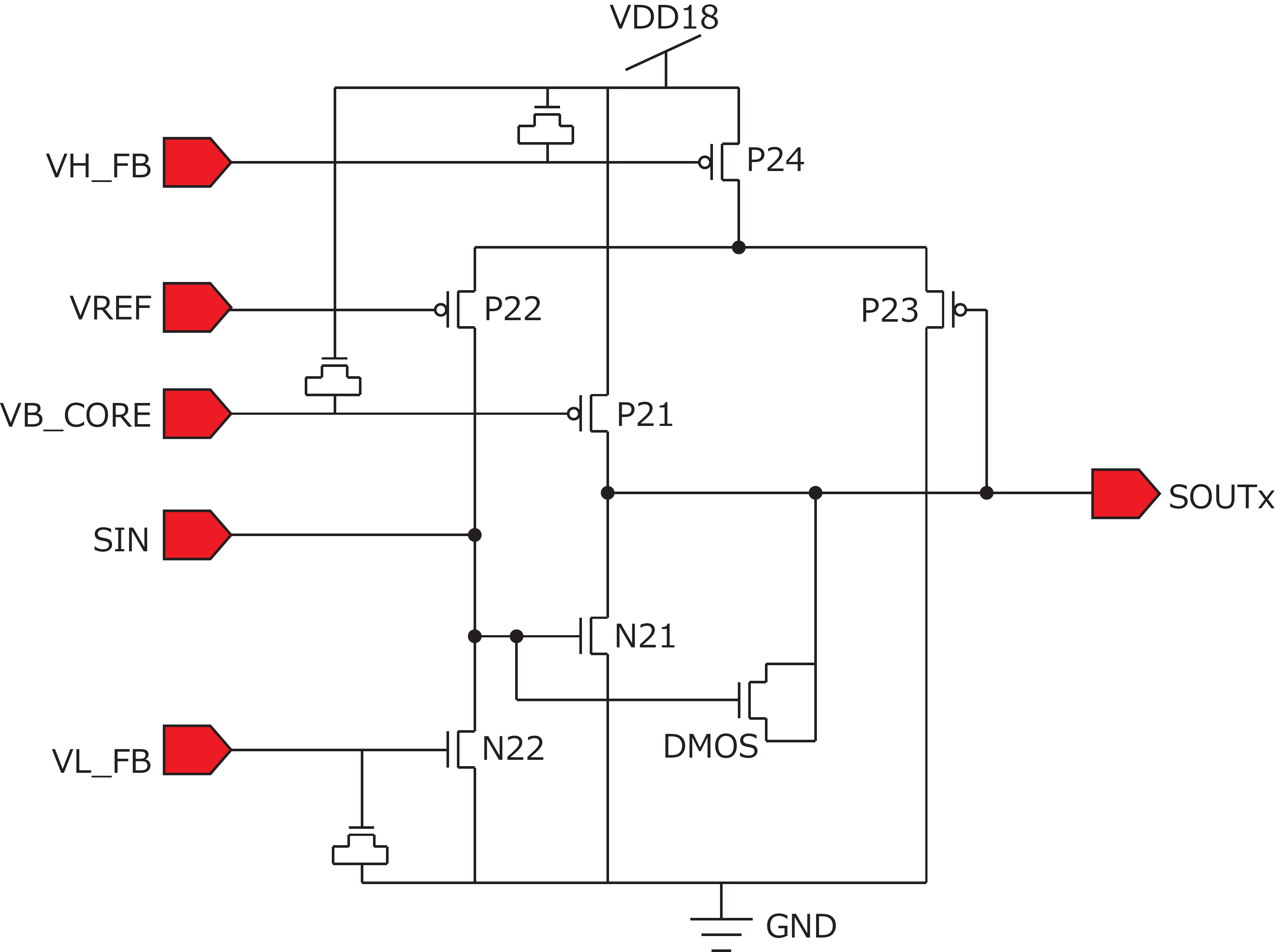}
\end{center}
\caption{The schematics of the shaper in ASD-TEG. The output signal from the amplifier is sent to SIN through a 35~fF capacitor. VREF voltage determines the input and output offsets. $\rm VB\_CORE$ ($\rm VH\_FB$ and $\rm VL\_FB$) voltage(s) generated from $\rm IIN\_CORE\_SHP$ ($\rm IIN\_FB\_SHP$) can adjust the gain (the time-width) of the shaper. The feedback current is larger in the shaper than in the amplifier.}
\label{fig:asdtegshp}
\end{minipage}
\end{figure}
\begin{figure}[htbp]
\begin{center}
\includegraphics[width=0.6\textwidth]{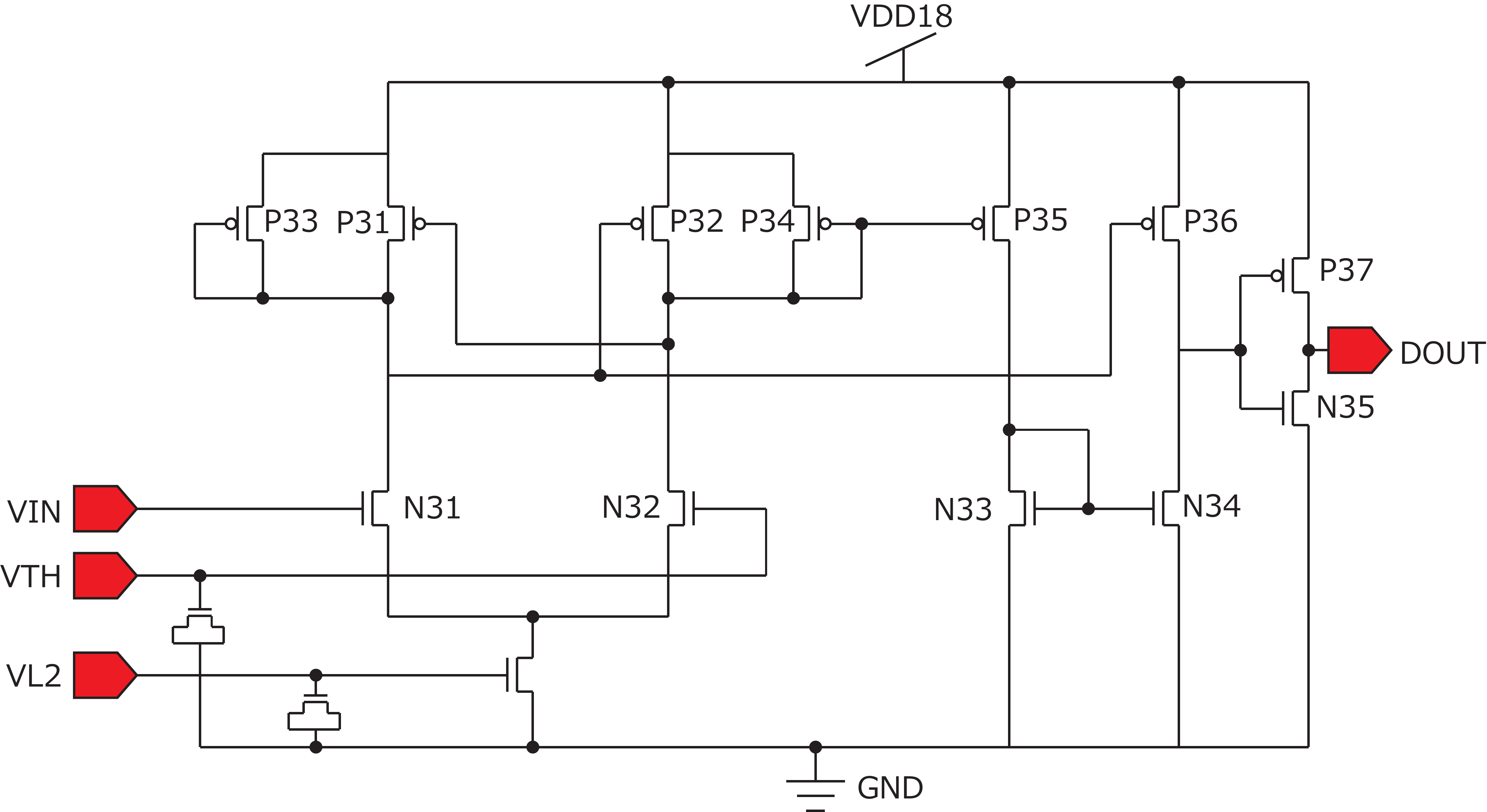}
\end{center}
\caption{The schematics of the discriminator in ASD-TEG. The discriminator has a hysteresis to avoid noise hits with two differential thresholds determined by a global threshold, VTH. The output signal is digital with 1.8~V for on-state and 0~V for off-state.}
\label{fig:asdtegdiscri}
\end{figure}

\clearpage

\section{Sensor-Circuit Cross-Talk Suppression}

\begin{figure}[htbp]
\begin{minipage}{0.48\hsize}
\begin{center}
\includegraphics[width=\textwidth]{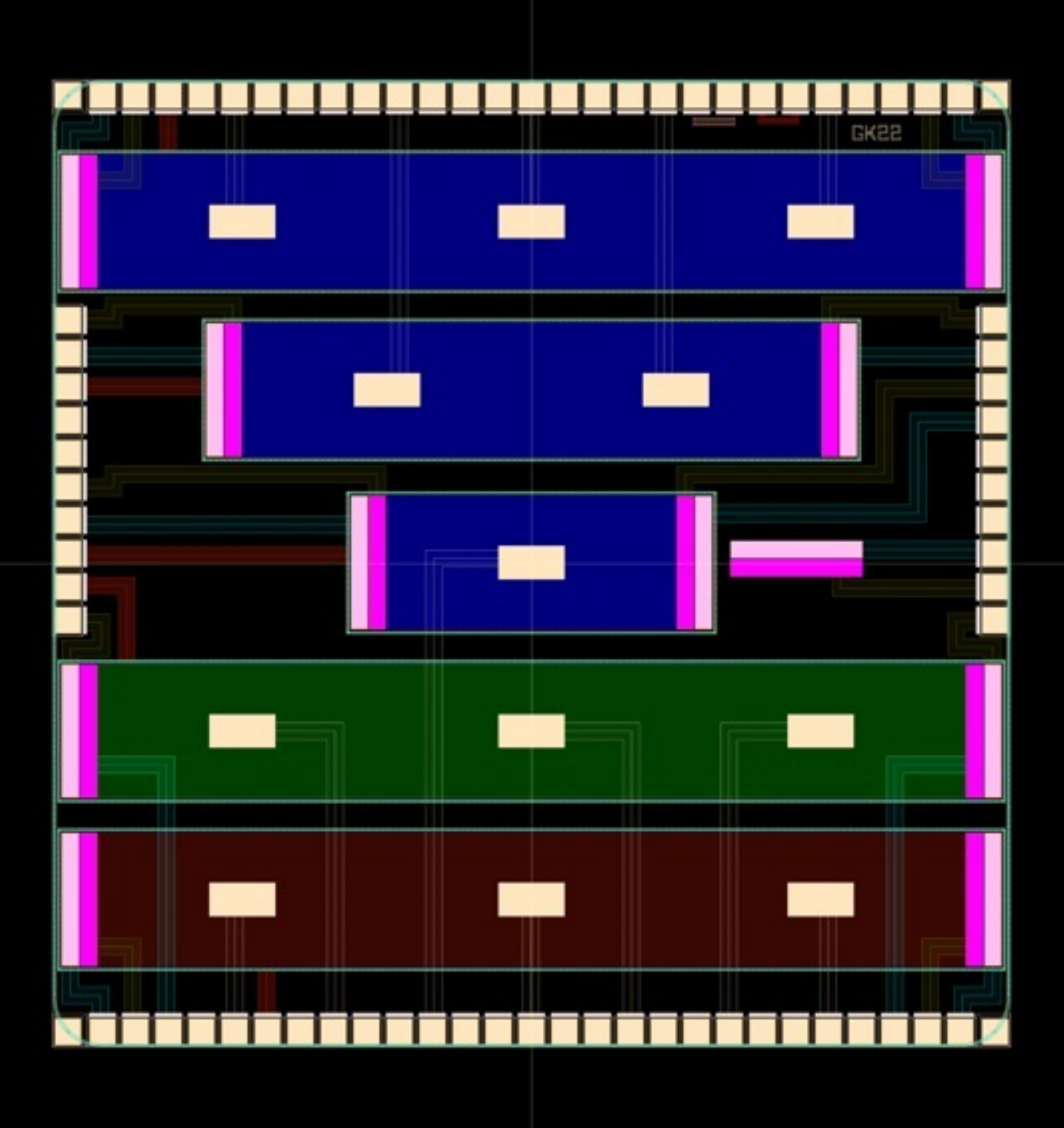}
\end{center}
\caption{The design of the CAPTEG having six capacitors. Three blue boxes are capacitors made between the BPW layer and SOI1 layer with difference sizes, the largest one (CAP1) and second largest (CAP2) being three times and twice larger than the smallest one (CAP3). The size of CAP3 is 875~$\rm\mu$m~$\rm\times$~400~$\rm\mu$m. }
\label{fig:capteg}
\end{minipage}
\hspace{0.04\hsize}
\begin{minipage}{0.48\hsize}
\begin{center}
\includegraphics[width=\textwidth]{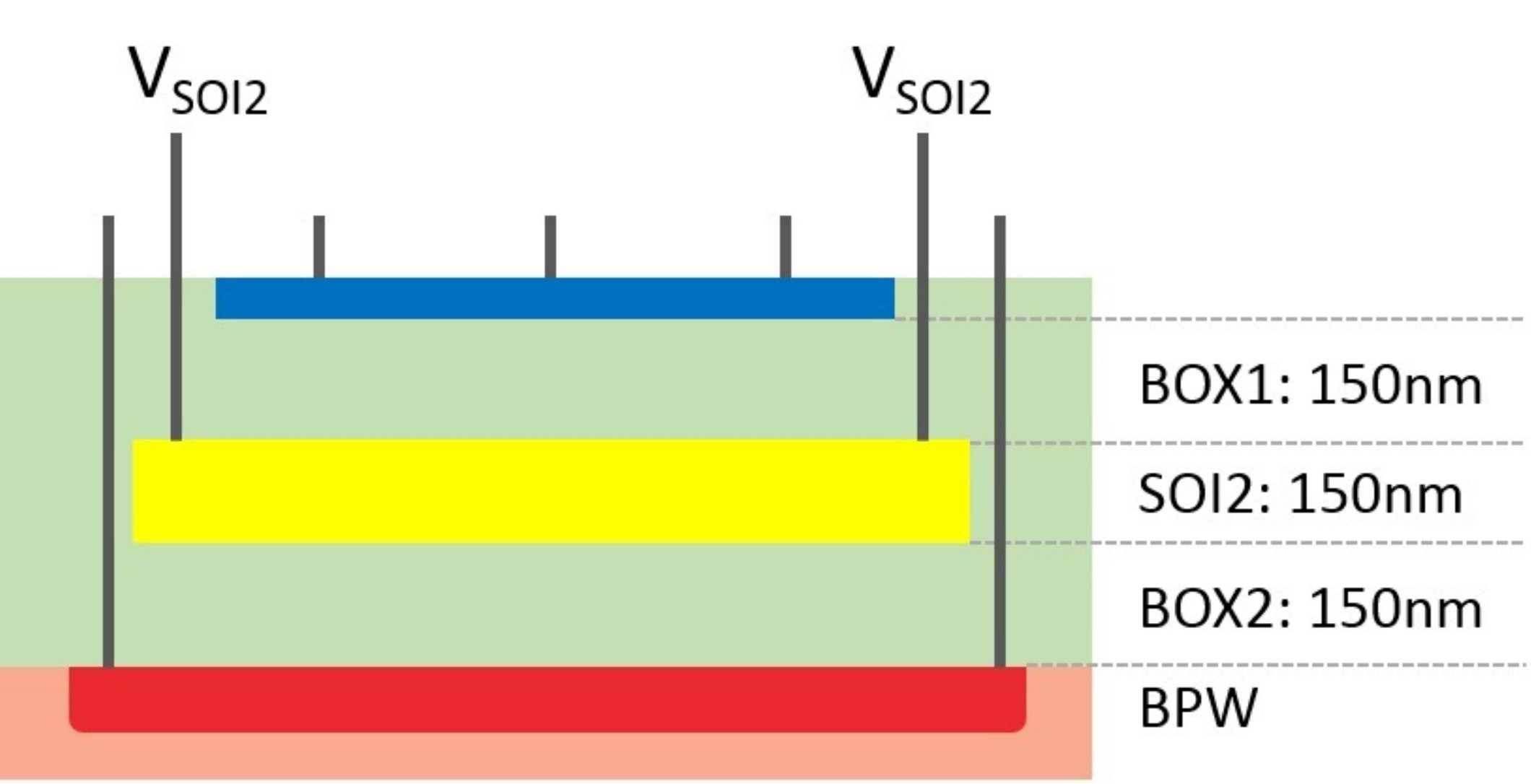}
\end{center}
\caption{The cross section of the CAPTEG, illustrating the capacitors constructed between the BPW layer and the SOI1 layer. The results using different readout pairs connected to each layer are identical and are averaged.}
\label{fig:captegxs}
\begin{center}
\includegraphics[width=\textwidth]{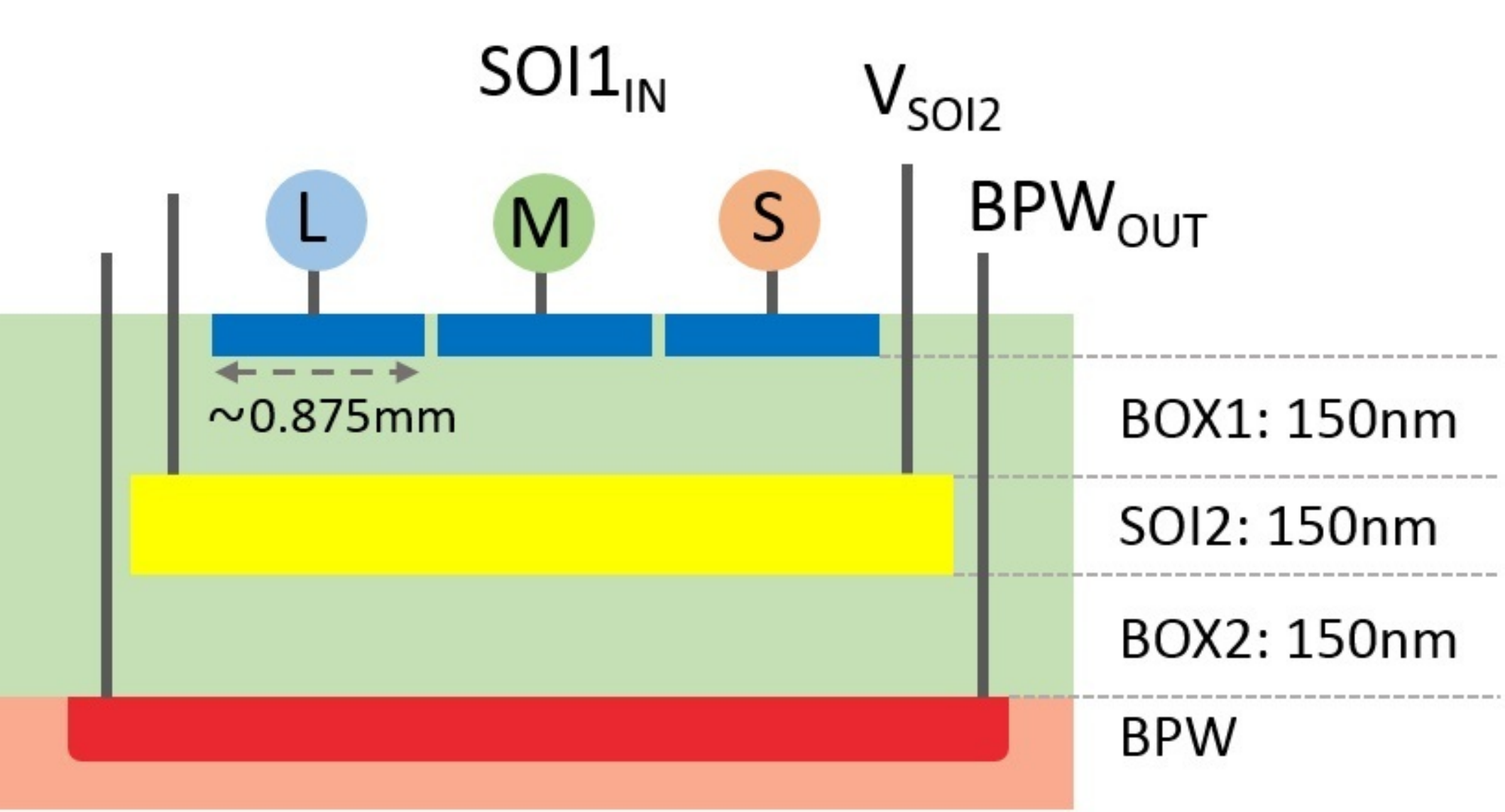}
\end{center}
\caption{The cross section of the XTALKTEG. Capacitances are named depending on the distance from the connection, S (=short), M (=medium), and L(=long). }
\label{fig:xtalk}
\end{minipage}
\end{figure}

\begin{figure}[htbp]
\begin{center}
\includegraphics[width=0.55\textwidth]{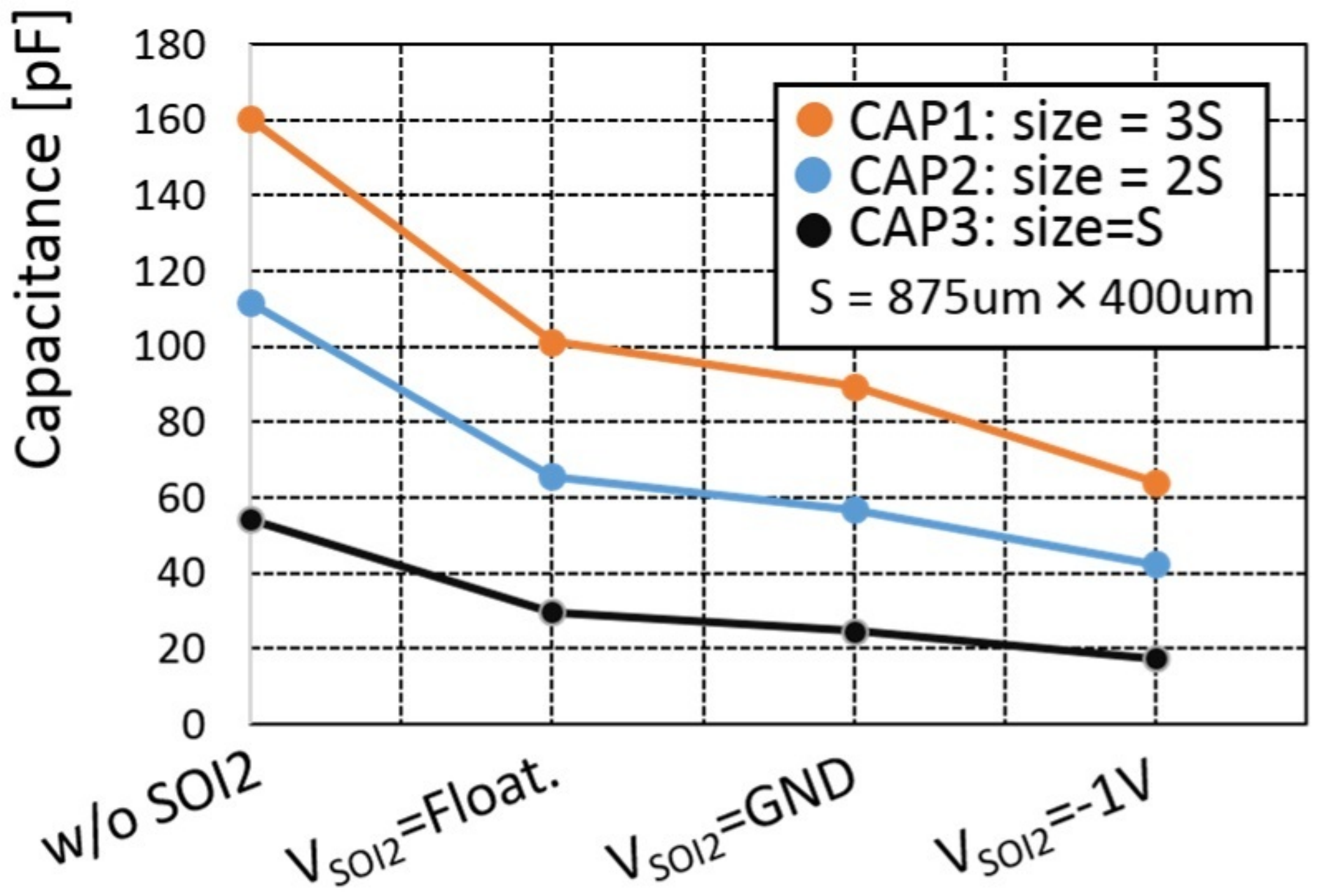}
\end{center}
\caption{The measured capacitances plotted for different SOI2 layer potential including floating. The values for w/o SOI2 are obtained for normal SOI device.}
\label{fig:captegresult}
\end{figure}

Two TEGs named CAPTEG and XTALKTEG are being evaluated for verifying sensor-circuit cross-talk suppression in DSOI. The CAPTEG is designed for a direct measurement of capacitances between the sensor layer and the circuit layer. The design of the CAPTEG is shown in Fig.~\ref{fig:capteg} with its cross section illustrated in Fig.~\ref{fig:captegxs}.
In CAPTEG, there are capacitances of three sizes made between the BPW layer as sensing-part and the SOI1 layer as the circuit-part (see Fig.~\ref{fig:captegxs}). The measured capacitance values are shown in Fig.~\ref{fig:captegresult} where different V$\rm _{SOI2}$ settings including floating are employed. The capacitance is reduced to about 50$\rm\%$ in DSOI (floating) compared to single SOI because the SOI2 layer separates the BOX-SOI1 capacitor into two capacitors connected in serial. Fixing the SOI2 potential at lower voltages (GNDed, -1~V, ...) reduced the capacitance further. Precise modelling of this suppression mechanism requires further studies and more detailed evaluation of the phenomena. 

\begin{figure}[htbp]
\begin{center}
\includegraphics[width=0.6\textwidth]{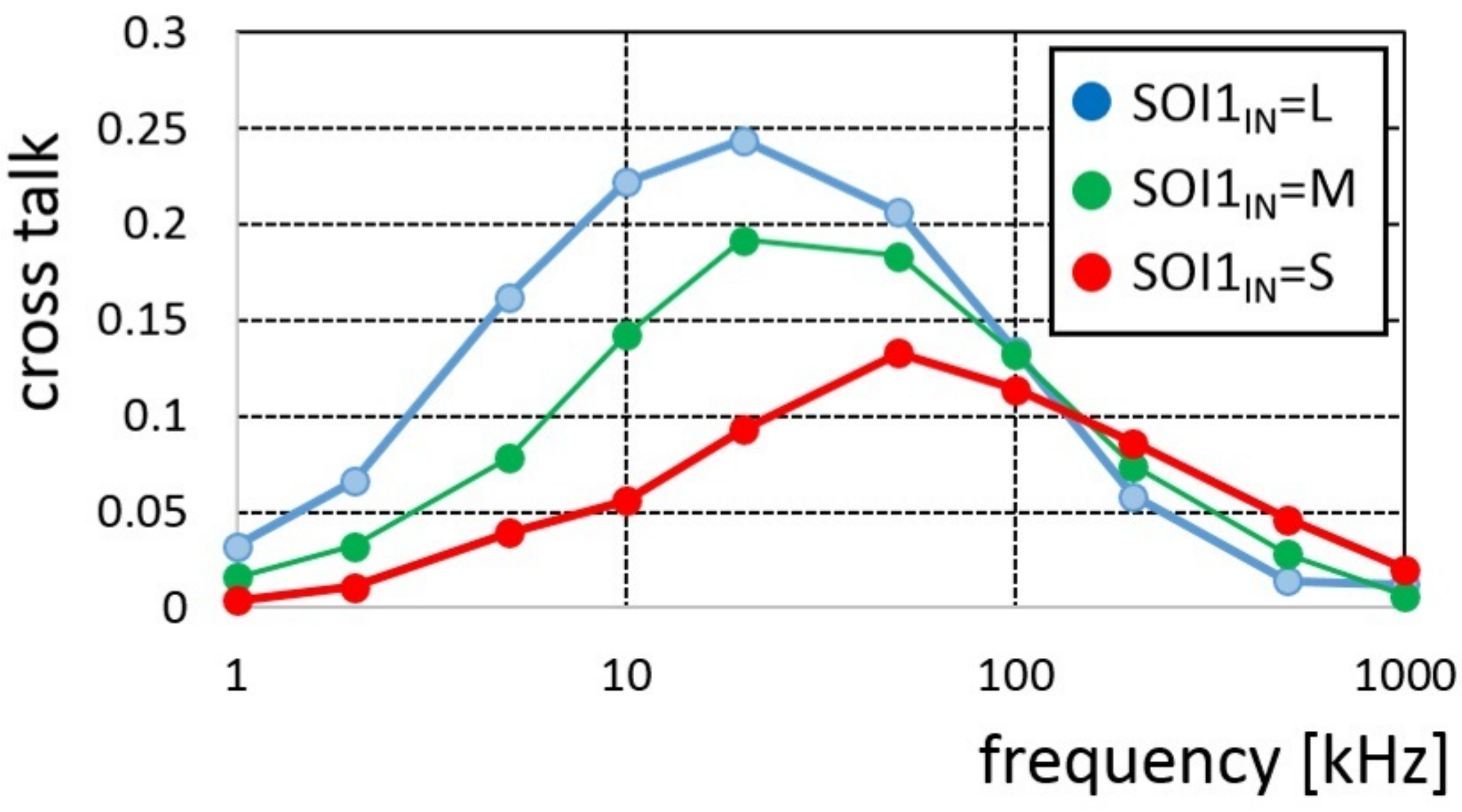}
\end{center}
\caption{The cross-talks magnitude as function of the signal frequency. The data are shown for three distances from the signal input and readout contacts. V$\rm _{SOI2}$ = -1~V. }
\label{fig:xtalkresult}
\end{figure}

For the case of XTALKTEG there are three capacitances arranged next to each other, where the SOI2 shielding effect of the cross-talk suppression can be evaluated taking into account of the finite resistance of the SOI2 layer. In Fig.~\ref{fig:xtalk}, the cross section of the XTALKTEG is shown. Because of the finite SOI2 resistance, the magnitude of cross-talk varies depending on distance between the readout position and the SOI2 connection. The cross-talk magnitude is defined as the ratio of the amplitude of the BPW layer output divided by that injected into the SOI1 layer. The input signal was 0.2~V sine-wave with various frequencies. The result is shown in Fig.~\ref{fig:xtalkresult}. The channel closer to the SOI2 connection has a better shielding performance as expected. The frequency dependence peaks at a certain frequency, and affects the sensor design in view of the cross-talk suppression. Therefore, although more detailed studies are needed, we can conclude that the suppression of the sensor-circuit cross-talk is effective in DSOI.

\clearpage

\section{Summary}
We are developing double-SOI pixel sensors. Successful TID compensation in a pixel ASD readout circuit has been demonstrated for radiation dose up to 100~kGy. The shaper width of less than 1 $\rm\mu$sec is maintained after 100~kGy. The discriminator also works properly after 100~kGy. 

The cross-talk suppression in double-SOI is being evaluated. The double-SOI is effective in reduction of the capacitance between the sensor node and the circuit electronics. The capacitance is reduced in double-SOI at V$\rm _{SOI2}$ = -1~V to 40$\rm\%$ compared to normal SOI. The cross-talk is substantially reduced if the distance from the SOI2 connection is shorter.

The obtained results further ensure that double-SOI sensors are applicable to future high-energy experiments such as at the BELLE-II experiment or at the ILC experiment. 

\Acknowledgements

The authors are grateful for fruitful collaboration with the Lapis Semiconductor Co. Ltd. The double SOI wafers have been realized through their excellence. This work was supported by JSPS KAKENHI Grand Number 25109006, by KEK Detector
Technology Project and also by VLSI Design and Education Center (VDEC), The University of
Tokyo, with the collaboration of the Cadence Corporation and Mentor Graphics Corporation.

\end{document}